\documentclass{emulateapj}

\shorttitle{The Hard-X-Ray Symbiotic RT Cru}
\shortauthors{Luna \& Sokoloski}

\begin{document}


\title{The Nature of the Hard-X-Ray Emitting Symbiotic Star RT Cru}


\author{G. J. M. Luna}
\affil{Instituto de Astronomia, Geof\'\i sica e Ciencias
  Atmosf\^ericas, Universidade de  
S\~ao Paulo, Rua do Mat\~ao 1226, Cid. Universitaria, S\~ao Paulo,
Brazil 05508-900} 
\email{gjmluna@astro.iag.usp.br}

\and

\author{J. L. Sokoloski}
\affil{Columbia Astrophysics Lab. 550 W120th St., 1027
  Pupin Hall, Columbia University, New York, New York 10027, USA}
\email{jeno@astro.columbia.edu}


\begin{abstract}
We describe $Chandra$ High-Energy Transmission Grating Spectrometer
observations of RT Cru, the first of a new sub-class of symbiotic
stars that appear to contain white dwarfs (WDs) capable of producing
hard X-ray emission out to greater than 50~keV.  
The production of such hard X-ray emission from the objects in this
sub-class (which also includes CD~$-57$~3057, T~CrB, and CH~Cyg)
challenges our understanding of accreting WDs.  We find that the 0.3
-- 8.0~keV X-ray spectrum of RT Cru emanates from an isobaric cooling
flow, as in the optically thin accretion-disk boundary layers of some
dwarf novae.  The parameters of the spectral fit confirm that the
compact accretor is a WD, and they 
are consistent with the WD being massive.  
We 
detect rapid, stochastic variability from the X-ray emission below 4
keV.  The combination of flickering variability and a cooling-flow
spectrum indicates that RT~Cru is likely powered by accretion through
a disk.  Whereas the cataclysmic variable stars with the hardest X-ray
emission are typically magnetic accretors with X-ray flux modulated at
the WD spin period, we find that the X-ray emission from RT~Cru is not
pulsed.  RT~Cru therefore shows no 
evidence for magnetically channeled accretion, consistent with our
interpretation that the $Chandra$ spectrum arises from an
accretion-disk boundary layer.

\end{abstract}




\section{Introduction}

Symbiotic stars are interacting binaries in which a hot, compact star
accretes from the wind of a red-giant companion.  Although a few
symbiotics contain neutron-star accretors \cite[e.g.,GX~1+4,
4U~1700+24, 4U~1954+31,
IGR~J16194-2810;][]{davidsen77,masetti02,galloway02,masetti07,masetti07b},
the accreting compact object is usually a white dwarf (WD).  Typical
binary separations are on the order of AU, with orbital periods on the
order of a few hundred days to a few decades
\citep[][]{kenyon,belczynski}.  Symbiotics can thus be thought of as
very large cousins of cataclysmic variables (CVs). The accretion rate
onto the WD appears to be high enough in most symbiotic systems that
accreted material is burned quasi-steadily in a shell on the WD
surface, producing a high UV luminosity \citep{jeno2001,orio}.
Although accretion disks are likely to exist around the WDs in some
symbiotics \citep{livio97,jeno2003}, there is little direct evidence
for these disks.  Finally, the red-giant wind produces a dense nebula
that surrounds the system.

Most symbiotic stars with detectable X-ray emission display soft, or
supersoft, thermal X-ray spectra.  As in the supersoft X-ray sources,
the lowest-energy X-rays could emanate directly from material burning
quasi-steadily on the WD surface \citep{jordan94,orio}.  Based on a
survey of symbiotics with $ROSAT$, \citet{murset97} proposed that
symbiotics be classified according to the hardness of their X-ray
spectra.  They labeled sources with supersoft spectra $\alpha$-types,
sources with the slightly harder spectra likely to arise from
collision of the red-giant and white-dwarf winds $\beta$-types, and
systems with the hardest spectra that might be indicative of neutron
stars $\gamma$-types.  Although more recent observations using the
broader X-ray coverage and greater sensitivity of $Chandra$ and
XMM-$Newton$ have shown that some symbiotics do not fit into the
simple $\alpha$/$\beta$/$\gamma$ classification scheme \cite[e.g.,
Z~And and $o$~Ceti;][]{zand,karovska2005}, most still appear to
produce primarily soft X-rays ($E < 3$ keV).

With the advent of the sensitive hard X-ray detectors on the $Swift$
and INTEGRAL satellites, a new picture has emerged.  Some symbiotic
stars can produce X-ray emission out to greater than 50~keV.  Such
hard X-ray emission has so far been detected from 4 symbiotics thought
to harbor WDs -- RT Cru \citep{integral,bird2007}, T CrB
\citep{tcrb,luna2007}, CH Cyg \citep{koji2007} and CD -57 3057
\citep{cd57,bird2007}.  Although the origin of this hard X-ray
emission is not known, there are some underlying similarities between
these hard X-ray emitting symbiotics.  Unlike most other symbiotic
stars, they display a high incidence of optical flickering.  They also
tend to have low optical line strengths, indicating that they are
often only ``weakly" symbiotic.  Both the visibility of the optical
flickering \cite[which in most symbiotics is overwhelmed by
reprocessed shell-burning emission;][]{soko03} and the weakness or
low-ionization-state of the optical lines suggest that quasi-steady
shell burning is not taking place in these objects, either because: 1)
the WD is more massive, or 2) the accretion rate is lower than in
other symbiotics.  Hard X-ray emission might therefore be a proxy for
high WD mass.  In fact, at least one of the hard-X-ray symbiotics,
T~CrB, is a recurrent nova and contains a high-mass WD
\citep{hachisu01}.  Finally, jet production appears to be more common
in flickering symbiotics, and one of the 4 hard X-ray symbiotics --
CH~Cyg -- regularly produces jets
\citep{taylor,karovska98,crocker2001, crocker2002}.  Other WDs in hard
X-ray symbiotics might thus also harbor jets.

\citet{dionisio94} classified RT~Cru as a symbiotic star based on its
optical spectrum.  They noted that the lack of strong high-ionization
emission lines and the very weak forbidden emission lines make the
optical spectrum similar to that of T~CrB.  They detected optical
flickering in the U band with a time scale of a few tens of minutes.
Except for GX~1+4 \citep{jabl97,chak97}, none of the neutron-star
containing symbiotic stars produce optical flickering
\citep[e.g.][]{jeno2001}, which is common in CVs, or show Balmer
or He~II emission lines \citep{masetti07b}.  The presence of
optical flickering and Balmer and He~II emission lines from RT Cru
\citep{dionisio94} suggests that it therefore contains an accreting WD
rather than a neutron star.  Reddening estimates from optical spectra
and infrared magnitudes \cite[coupled with the assumption that the
radius of the M5 III red giant is about 0.5~AU;][]{vanbelle99} suggest
that RT Cru is between 1.5 and 2~kpc away (J. Miko{\l}ajewska, {\em
private communication}).

In 2003 and 2004, the IBIS instrument on board INTEGRAL detected hard
X-ray emission extending out to $\sim$100~keV from the source
IGR~J12349-6434, which \cite{integral} found to have a 20-60~keV flux
density of $\sim$3~mCrab.  \cite{masetti} suggested an association
between IGR~J12349-6434 and RT~Cru, which observations with the
$Swift$ satellite later confirmed \citep{swift}.  The long-term
optical light curve from the AAVSO indicates that at the time of the
INTEGRAL observations, RT Cru was in an optical bright state; it
brightened from 13.5 to 11.5 mag sometime between 1998 and 2000.
Between 2000 and 2005, the optical brightness slowly decreased to
approximately 12.1 mag.  The short (4.7~ks) Swift observation of 2005
August showed that between 2003 and 2005, the hard X-ray flux also
decreased.

In this paper, we describe $Chandra$ High Energy Transmission Grating
(HETG) observations of RT~Cru, the first member of a new class of hard
X-ray emitting symbiotic WDs.  We detail the observations and data
reduction in \S\ref{sec:obs} and the results from spectral and timing
analyses in \S\ref{sec:results}.  In \S\ref{sec:disc}, we discuss our
interpretation of the observations, which confirm that the accreting
compact object in RT~Cru is a WD and provide some of the most direct
evidence to date for an accretion disk around a wind-fed WD in a
symbiotic system.  In this section, we also discuss the implications
of a system that can accelerate particles to relativistic speeds and
produce X-ray emission out to greater than 50~keV being powered by an
accreting WD.  We summarize our conclusions in \S\ref{sec:conclusion}.

\section{Observations and Data Reduction} \label{sec:obs}

\begin{figure}[ht]
\includegraphics[width=6cm,angle=-90]{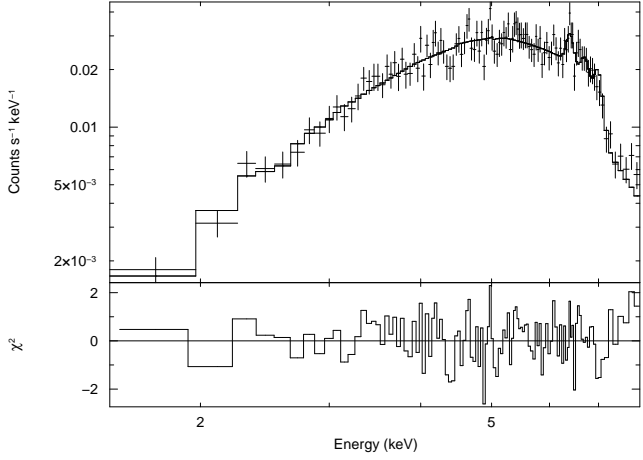} 
\caption{Undispersed (zeroth order) spectrum. The top panel shows the
  spectrum with the absorbed, isobaric cooling-flow model
  over-plotted.  The bottom panel shows residuals with respect to this
  model (in units of $\chi^2$, where $\chi^2$ is shorthand for the
  difference between the data and the model, squared, divided by the
  variance, with the sign of the difference between the data and the
  model).
}
\label{zero_order}
\end{figure}

On 2005 October 19, the $Chandra$ X-ray Observatory performed a
50.1~ks Director's Discretionary Time observation of RT Cru using the
HETG \citep{can05} and the ACIS-S detector (ObsId 7186, start time
10:21:12 UT).  We requested the DDT observation to attempt to catch
RT Cru in the optical bright state that appeared to be associated with
hard X-ray emission.  We used the HETG instrument because the 
Swift XRT observation of 2005 August 
hinted at several possible emission-line complexes.  The data were
collected in Timed Exposure mode, in which the CCD chips were read out
every 2.54~s.  The data were telemetered back to earth in Faint mode,
which conveys photon arrival times, event amplitudes, and additional
information for evaluating the validity of each event.  We reduced the
data according to standard procedures using the software package
CIAO~3.3\footnote{Chandra Interactive Analysis of Observations (CIAO),
http://cxc.harvard.edu/ciao/.}.  We extracted a spectrum from the
undispersed light (the zeroth-order spot, which fell on the S3,
back-illuminated chip) using a circular extraction region with a
radius of 6\arcsec\, centered on the source coordinates: $\alpha =
12$h~34m~43.74s and $\delta =-64^{\circ}$ 33\arcmin 56.0\arcsec.  To
obtain the background for the zeroth-order light, we extracted photons
from a source-free circular region on the same CCD.  We grouped the
spectrum, which is shown in Fig.~\ref{zero_order}, to have at least 50
counts per bin. The average zeroth-order source count rate was 0.11
c~s$^{-1}$.

For the dispersed light from both of the HETG sets of gratings -- the
Medium Energy Grating (MEG) and the High Energy Grating (HEG) -- we
extracted spectra from each of the $m=\pm 1$, $\pm 2$, and $\pm 3$
orders individually (using the CIAO software tool dmtype2split).  To
obtain the background for the dispersed light, we extracted counts
from rectangular regions on either side of the spectral image.  The
count rate in the HEG and MEG $m\pm 1$ orders was 0.042 c~s$^{-1}$ and
0.034 c~s$^{-1}$, respectively.  Although the dispersed spectral
orders contained too few counts to produce a high
signal-to-noise-ratio spectrum of lines spanning the full energy range
of the instrument, the combined $m = \pm 1$ spectrum (grouped at twice
the full width at half maximum of 0.012\AA) provided good-quality data
in the region around the Fe K$\alpha$ emission-line complex.  We
therefore used the the zeroth-order spectrum for continuum fitting and
the dispersed ($m= \pm 1$) spectra primarily for analysis of the Fe
lines.  The HEG and MEG $m =\pm$2 and $\pm 3$ spectra helped confirm
the Fe line identifications.  For spectral fitting of both the zeroth
and higher order photons, we used the standard software packages Xspec
\citep{xspec} v12.3.0 and ISIS \citep{isis}.  The background
contributed less than 1\% of the total extracted dispersed and
undispersed light.

We generated light curves in the energy bands 0.3--4.0~keV and
4.0--8.0~keV by extracting counts (with CIAO) from a region containing
the zeroth-order spot and the $m=\pm 1$ dispersed orders of both the
HEG and MEG.  Since we estimated there to be only $\sim$70 background
counts in this extraction region during the course of the observation
(compared to more than 9000 source counts), we did not
background-subtract the light curves.

At high count rates, two or more photons can arrive close enough
together in time that they appear to be a single event.  This
``pileup'' phenomenon can cause a spectrum to become distorted and the
count rate to be reduced.  To confirm that the zeroth order spectrum
was not affected by pileup, we divided the number of counts at the
peak of the point spread function of the undispersed light by the
number of 2.54-s frames in the observation to obtain an upper limit on
the number of counts per pixel per frame.  The resulting 0.08 counts
per pixel per frame is well below the 1 count per pixel per frame
where pileup can become important \citep{harris}.  Moreover, the
average count rate in the zeroth order was significantly less than one
count per frame time, indicating that the light curve was not
significantly distorted by the saturation that can occur at higher
count rates (i.e., higher pileup fractions).  The pileup fraction in
the higher-order spectrum was negligible.

\section{Analysis and Results} \label{sec:results}

\subsection{Spectral Analysis}

To model the X-ray spectrum, we first consider simple,
single-component continuum models.  We fitted these models to the
binned 0.3 -- 8.0~keV zeroth-order spectrum (above 8.0~keV, the noise
rises and the quantum efficiency drops sharply).
Absorbed single-component emission models such as a thermal
plasma, powerlaw, or blackbody (plus Gaussian lines) do not produce
acceptable fits.  Even if we include complex absorption, such as an
absorber that only partially covers the source, a powerlaw
distribution of absorbers \cite[as seen in some magnetic CVs;
e.g.,][]{done98}, or a``warm" ionized absorber, single-component
emission models still do not produce acceptable fits.

Including an additional broad-band emission component improves the
fitting results.  The spectrum is formally well fitted with a highly
absorbed, optically thin thermal plasma (Mekal model in Xspec), plus a
moderately absorbed non-thermal powerlaw component.  Since there is
some degeneracy between the amount of absorption and the powerlaw
index, we determine the powerlaw index by fixing the plasma
temperature, $T$, and fitting the spectrum above 4~keV, where
absorption is relatively unimportant.  The resulting photon index is
$\Gamma = 1.05_{0.78}^{1.29}$, where $\Gamma$ is given by $dF_{N}/dE =
K (E/E_{0})^{-\Gamma}$, $dF_{N}/dE$ is the photon flux density, $E_0$
is 1~keV, $K$ is the normalization constant, and the superscripts and
subscripts are 90\% confidence upper and lower limits, respectively.
The photon index is not sensitive to the value to which we fix $T$.
We determined the remaining model parameters by fixing $\Gamma$ to 1.1
and letting the other parameters vary.  The thermal plasma has $kT =
8.6_{6.1}^{12.4}$~keV and an absorbing column $n_H = 9.3_{7.2}^{12.5}
\times 10^{23}\, {\rm cm}^{-2}$.  The powerlaw component has an
absorbing column of $n_H = 7.3_{6.3}^{8.7} \times 10^{22}\, {\rm
cm}^{-2}$ and a normalization $K = 1.2_{0.98}^{2.9}\times
10^{-3}\,{\rm photons}\,{\rm cm}^{-2}\,{\rm s}^{-1}\,{\rm keV}^{-1}$.
To obtain an acceptable fit to the line emission, the model required
abundances of about 0.3 times solar \cite[using the abundances
of][]{abund}.  Because of the large column density absorbing the
thermal emission in this model, the powerlaw component dominates below
approximately 4~keV, and the thermal emission dominates above 4~keV.
We also obtained a formally acceptable fit with a highly absorbed
powerlaw and a moderately absorbed thermal plasma.  In that case, the
thermal plasma dominated the low-energy portion of the spectrum, and
the powerlaw dominated the high-energy part of the spectrum.

Finally, we consider the isobaric cooling-flow model that has worked
well for both magnetic and non-magnetic CVs.  In this model
\cite[Mkcflow;][]{mkcflow}, gas is assumed to radiatively cool from a
high post-shock temperature under conditions of constant pressure.
Such an isobaric cooling flow has a differential emission-measure
distribution that is a flat function of temperature.  The gas is also
assumed to be optically thin.  This cooling flow model provides a good
fit to the data.  We find that the initial post-shock temperature is
quite high.  Although our 0.3--8.0~keV spectrum does not allow a
high-confidence determination of this parameter, the formal 90\%
confidence lower limit for the initial temperature is $kT_{max} =
55$~keV.  The minimum cooling-flow temperature is consistent with the
smallest value allowed by the Xspec Mkcflow model ($kT_{min} =
80.8$~eV), indicating that the gas does indeed remain optically thin
as it cools.  Allowing the differential emission measure to have a
powerlaw distribution (with the Cemkl model in Xspec) did not improve
the fit.  As with the two-component continuum model, the cooling-flow
model requires significant absorption.  The abundances may also be
sub-solar \cite[0.3 times solar, using the abundances of][]{abund}.
The best absorption model consists of both a photoelectric
absorber that fully covers the source and another that only partially
covers it.  Table~\ref{tab1} lists the best-fit parameters for this
model, which, as we discuss in \S\ref{sec:disc}, we believe provides
the best description of the $Chandra$ spectrum of RT~Cru.

\begin{figure}[ht]
\includegraphics*[width=6cm, angle=-90]{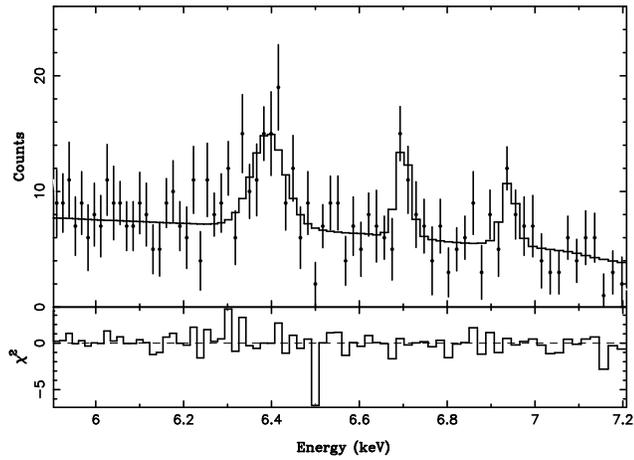}
\caption{The iron-line complex from the combined HEG and MEG
first-order ($m = \pm$1) spectrum.  The best fit model of a powerlaw plus
three Gaussian emission lines is over-plotted.  The  bottom panel
shows the residuals, in the same units as Fig.~\ref{zero_order}.}
\label{felines}
\end{figure}

In the first-order spectrum, we detect the iron-line complex spanning
roughly 6.4 -- 7.0~keV.  Fig.~\ref{felines} shows the region around
the iron-line complex in the combined MEG and HEG first-order ($m=
\pm$1) spectrum.  Because of the large absorption and the resulting
low count rate at low energies, we are not sensitive to lines such as
OVIII ($\sim 19$~\AA) and Ne X ($\sim 12$~\AA) that have been seen in
HETG observations of some other accreting WDs
\cite[e.g.,][]{pandel,mukai2003}.  For the Fe lines, we use a simple
powerlaw to establish a continuum level and three Gaussian profiles to
fit the Fe K$\alpha$, H-like Fe, and He-like Fe lines,
respectively.  To avoid the possible introduction of errors from
misalignment of the HEG and MEG spectra, we use only the combined HEG
first-order ($m = \pm 1$) spectrum for computation of the equivalent
widths (EWs).  Table~\ref{tab2} lists the line-center energies and
EWs. Although we do not have enough counts to use the recombination,
intercombination, and forbidden components of the H- or He-like Fe
lines as density diagnostics, the observed line strengths and EWs
confirm that the source is surrounded by a large amount of neutral
material and that the abundances might be slightly sub-solar.  The Fe
K$\alpha$ EW of 108~eV is consistent with that expected for a source
inside a cloud of cold material with the N$_H$ of $\sim$ 10$^{23}$
cm$^{-2}$ \citep{inoue}, as we found from the continuum fitting.

\subsection{Timing Analysis}

\begin{figure}[t]
\includegraphics[width=8cm]{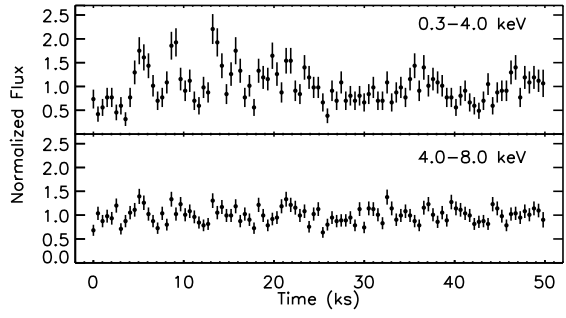}
\caption{$Chandra$ light curves for RT~Cru, with a bin size of
  508.28~s.  The light curves include the undispersed light as well as
  the counts from the HEG and MEG $m=\pm 1$ orders.  The top and
  bottom panels show the flux as a function of time in the energy
  ranges 0.3--4.0~keV and 4.0--8.0~keV, respectively.  The
  0.3--4.0~keV emission is clearly variable on time scales of minutes to hours.}
\label{fig:lcs}
\end{figure}

Examining 
time series binned at 508.208~s and 4065.664~s (i.e., 200
and 1600 times the frame time, respectively), we detected significant
aperiodic, flickering-type variations on time scales of minutes to
hours in the 0.3--4.0~keV emission.  Figure~\ref{fig:lcs} 
shows the 508-s 
binned time series (light curves)
in the energy ranges 0.3--4.0~keV and 4.0--8.0~keV.  The
fractional amplitude of the stochastic variations appears to be
largest in the 0.3--4.0~keV energy range.  In the 508-s binned
0.3--4.0~keV time series, the ratio of measured fractional rms
variation, $s$, to that expected from Poisson fluctuations alone,
$s_{exp}$, is 1.96 ($s=36.6$\% and $s_{exp}=18.7$\%).  We detect the
0.3--4.0~keV variability with even greater statistical significance in
the 4065-s binned time series; the ratio $s/s_{exp}$ in this case is
3.35 ($s=22.1$\% and $s_{exp}=6.6$\%).  In the 4.0--8.0~keV energy
range, the 508-s binned time series has $s/s_{exp} = 1.36$ (where
$s=16.6$\% and $s_{exp}=12.2$\%), and the 4065-s binned time series
has $s/s_{exp} = 1.42$ ($s=6.1$\% and $s_{exp}=4.3$\%).

We do not detect any periodic flux modulations.  We are
theoretically sensitive to an oscillation with fractional amplitude:
\begin{eqnarray} \label{eqn:sens}
A & \approx & 2 \left(C_{tot} \alpha
\right)^{-1/2} \left (\ln 
\left[\frac{1-\delta}{ 1-(1-\epsilon)^{1/n_{freq}}}
  \right]\right)^{1/2} \\
 & = & 0.08 \left( \frac{C_{tot}}{9,400} \right)^{-1/2} \left(
\frac{\alpha}{0.77} \right)^{-1/2},
\end{eqnarray}
where $C_{tot}$ is the total number of counts in the observation
(ignoring the small number of background counts, which have a
negligible effect), $\delta$ and $\epsilon$ are small numbers related
to the chance that a noise power in the power spectrum will exceed the
detection threshold (both taken to be 0.05), $n_{freq}$ is the number
of frequencies searched ($n_{freq} = 1644$), and $\alpha$ has an
average value of 0.77 and depends upon the location of the signal
frequency in the frequency bin \cite[see,
e.g.,][]{vanderklis89,soko99}.  We are therefore sensitive to
oscillations with fractional amplitudes of $\sim 8$\% in regions of
the power spectrum dominated by white noise, which in this case
consisted of frequencies greater than $\sim 1.4$~mHz.  In this
analysis, we binned the time series in 15-s bins (i.e., 6 times the
frame time).  We were therefore sensitive to oscillations with periods
as short as 30~s, and most sensitive to oscillations with periods
between 30~s and 12~m.

Consistent with the presence of flickering in the light curves, the
power spectrum rises at frequencies below 1.4 mHz.  At these low
frequencies, the power spectrum has a powerlaw index (i.e., slope on a
log-log plot) of about $-1$.  This ``$1/f$-noise'' at low frequencies
reduces our sensitivity to oscillatory signals with periods greater
than approximately 12~m.  Since we expect the minimum oscillation
amplitude to which we are sensitive to increase roughly as the square
root of the rising average broadband power as we go to lower
frequencies (i.e., longer periods), the oscillation amplitude required
for detection increases gradually from $\gtrsim$8\% to $\gtrsim$15\%
as we move from periods of 12~m to 1~hr.  Taking into account the
underlying broadband power, as well as the number of frequency bins
searched, we did not detect any statistically significant oscillations
in any portion of the power spectrum.


\section{Discussion} \label{sec:disc}

\subsection{Interpretation of the $Chandra$ Observations}

To estimate the radius of the accreting compact object, we take the
unabsorbed 0.3--8.0~keV X-ray luminosity, $L_X$, to be either
approximately equal to, or a rough lower limit to, the emission from
an accretion-disk boundary layer (we justify the assumption that the
$Chandra$ X-ray emission emanates from a boundary layer in the
paragraphs that follow).  Comparing this luminosity with that expected
from accretion, $(1/2) (G M \dot{M} / R) \gtrsim L_X$,
the radius of the accreting compact object is:
\begin{equation}
R \lesssim 3.2 \times 10^{8}\, {\rm cm} \left( \frac{M}{1.3 M_\odot} \right)
\left( \frac{\dot{M}}{1.8 \times 10^{-9} M_\odot\,{\rm yr}^{-1}} \right),
\end{equation}
where $R$ and $M$ are the radius and mass of the accretor,
respectively, and $\dot{M}$ is the rate of accretion through the
boundary layer.  The radius is that of a WD.  The $Chandra$ X-ray
spectrum therefore confirms that the compact object 
is a WD.  

To determine whether the $Chandra$-band X-ray emission is indeed from
an accretion-disk boundary layer, we consider the rapid variability.
Rapid flickering typically emanates from an accretion region close to
a compact object.  Our detection of flickering therefore suggests that
the X-ray emission detected from RT~Cru by $Chandra$ is powered by
accretion.  This accretion could proceed via a wind-fed accretion
disk, magnetic accretion columns, or Bondi-Hoyle type direct-impact of
the accreting material onto the WD.  While the two-component (thermal
plasma plus powerlaw) model provides a formally acceptable fit to the
data, it is difficult to construct an interpretation of this model
that is consistent with the rapid flickering from accretion onto a WD.
The isobaric cooling-flow spectral model, on the other hand, 1) provides
a good fit to the data, 2) has been successfully applied to both the
boundary layer emission from non-magnetic CVs
\cite[e.g.,][]{mukai2003,pandel} and the accretion columns of magnetic
CVs \cite[e.g.,][]{cropper}, and 3) provides a natural context for the
flickering from accretion.

Most CVs with X-ray emission as hard as that which INTEGRAL and
$Swift$/BAT have detected from RT~Cru have magnetic fields strong
enough to channel the accretion flow into accretion columns ($B \sim
10^5$ -- $10^6$~G, where $B$ is the magnetic field strength at the WD
surface).  Of the 8 CVs detected at energies greater than $\sim
50$~keV with $Swift$/BAT, all but SS Cyg are likely magnetic accretors
\citep{barlow06}.  In these systems, the hard X-rays come from hot gas
behind the stand-off shock in the accretion column.  Magnetic CVs
typically have X-ray oscillations with pulsation amplitudes of tens of
percent at the WD spin period, which is usually less than an hour
\citep[][]{war95}.  Since typical symbiotic-star accretion rates are
higher than typical CV accretion rates, if the WD in RT~Cru was
strongly magnetic and in spin equilibrium, the spin period would
probably be either comparable to or faster than those in CVs.  Given
our sensitivity to oscillations with periods less than an hour, we
therefore should have detected a spin modulation if RT~Cru was
magnetic.  In fact, the power spectrum has no statistically
significant peaks.  We conclude that RT~Cru is probably not a magnetic
accretor.  We thus favor the picture in which the X-ray emission from
RT~Cru detected by $Chandra$ is from a cooling flow in an
accretion-disk boundary layer.

\subsection{Implications}

The parameters of the cooling-flow fit to the boundary-layer emission
provide the accretion rate as well as information about the WD.  From
the normalization parameter of the cooling-flow model, the accretion
rate onto the WD is $\dot{M}= 1.8 \times 10^{-9}$
M$_{\odot}$~yr$^{-1}\,(d / 2\, {\rm kpc})^2$ (see Table~\ref{tab1}).
Since quasi-steady nuclear burning is frequently present on the WD
surface, symbiotic stars should have accretion rates that are on
average higher than those in CVs.  The accretion rate we have found
for RT~Cru is consistent with this picture.  It is also, however, just
low enough that we expect the boundary layer to remain optically thin;
\citet{narayan93} find that for a $1\,M_\odot$ WD, the boundary layer
remains optically thin for accretion rates below $3 \times 10^{-9}\,
M_\odot\, {\rm yr}^{-1}$.

The parameters of the cooling-flow fit also indicate that the WD
radius is small, suggesting that the WD could be quite massive.  Taken
at face value, the radius constraint would imply that the WD mass is
at least 1.3~$M_\odot$.  The high upper cooling-flow temperature,
$kT_{max} \ge 55$~keV (see Table~\ref{tab1}), supports the conclusion
that the accretor is a massive WD.  The relationship between
$kT_{max}$ and WD mass is due to the fact that that the Kepler
velocity ($v_{\rm K}^2 = G M / R$) is greater in the deep potential
well of a more massive WD.  Since the boundary-layer material is shock
heated, and the initial post-shock temperature ($T_{max}$) is
proportional to velocity squared, $T_{max}$ increases with WD mass.
Alternatively, if we equate the amount of energy available per
particle, $(1/2) \mu m_p v_{\rm K}^2$, where $\mu$ is the mean
molecular weight and $m_p$ is the mass of a proton, with the energy
released per particle in an isobaric cooling flow, $(5/2) k T_{max}$
\citep{pandel}, we see that $ k T_{max} \propto v_{\rm K}^2 \propto
GM/R$.  Although the determination of $kT_{max}$ from X-ray emission
below 8~keV is highly uncertain, we can still ask what such a high
$kT_{max}$ would imply if it is confirmed by an instrument with
greater high-energy sensitivity (such as SUZAKU).  For their sample of
9 non-magnetic CVs, \citet{pandel} found the upper cooling-flow
temperature was roughly consistent with the expected $kT_{max} =
(3/5)\, kT_{vir}$, where $T_{vir}$ is the virial temperature (defined
by $(3/2)kT_{vir} = (1/2) \mu m_p v_{\rm K}^2 $).  Using the WD
mass-radius relationship of \citet{hkbook}, a maximum cooling-flow
temperature of $kT_{max} > 55$~keV implies $M \gtrsim 1.3\,M_\odot$.

The high absorbing columns for both the partial-covering and fully
covering absorber, as well as the covering factor of $>$0.7 for the
partial-covering absorber (see Table~\ref{tab1}), indicate that the
X-ray source is highly obscured at this epoch.  Since the X-ray
emission region is small, the absorber that only partially covers the
X-ray source must also be small. A possible source of this partially
covering absorber is an accretion structure such as an accretion disk
seen almost edge-on \cite[as in OY Car;][]{pandel}. We assume that the
fully covering component of the absorption comprises both interstellar
absorption (1.1$\times$10$^{22}$cm$^{-2}$ from NASA/IPAC IRSA) and
intrinsic absorption.  The column density of this absorber is probably
high because the WD orbits within the strong, dense stellar wind from
the red giant.  RT~Cru has an orbital period of $\sim 450$~d
(J. Miko{\l}ajewska, {\em private communication}), and therefore a
binary separation on the order of an AU.  This separation puts the WD
well within the dense region of the red-giant wind.  Month-time-scale
variations in the absorption (not correlated with the orbital period)
have been detected by $Swift$ \citep{kennea07}, suggesting that either
the red-giant wind is clumpy, that the mass loss rate in the wind of
the red giant is variable, or the accretion structure or structures
partially blocking the WD boundary layer must be unstable.

\subsection{The Nature of RT Cru}

As shown by the INTEGRAL detection in 2003-2004 \citep{integral},
RT~Cru can at times produce X-ray emission out to greater than 60~keV.
From the broad-band flux densities reported by \citet{integral}, the
hard X-ray spectrum in 2003 - 2004 appears consistent with a powerlaw
with photon index $\Gamma = 2.7$ (energy index $\alpha = 1.7$), and
the 16-100~keV luminosity at that time was approximately
$10\,L_\odot\,(d/2{\rm kpc})^2$.  By 2005, however, the 16-100~keV
luminosity had dropped, and the hard X-ray spectrum was closer to
thermal \citep{kennea07}.  To better appreciate the properties of the
unusual accreting WD in RT~Cru, we briefly explore the possible
sources of the powerlaw hard X-ray emission observed by INTEGRAL in
2003-2004.  In particular, we consider direct synchrotron emission,
inverse-Compton (IC) scattering from a thermal distribution of
electrons, and IC scattering from a non-thermal distribution of
electrons.  IC scattering from a non-thermal distribution of electrons
turns out to be the most likely option.

One way to generate a powerlaw energy spectrum is with synchrotron
emission from a powerlaw distribution of relativistic electrons moving
in a magnetic field.  Most of the synchrotron emission from an
electron with Lorentz factor $\gamma$ is emitted at a frequency
$\nu_{sync} = (0.3/2 \pi) (3 \sin \alpha / 2) \gamma_{sync}^2 (q B /
m_e c)$, where $\alpha$ is the pitch angle, $q$ is the electron
charge, $B$ is the magnetic field strength, $m_e$ is the electron
mass, and $c$ is the speed of light \citep{rlbook}.  
Since we did not detect X-ray pulsations from RT Cru, the magnetic
field at the surface of the WD is probably not strong enough to
disrupt the accretion flow, and therefore less than $\sim 10^4$~G.
Thus, even near the
surface of the WD, 
Lorentz factors $\gamma_{sync}$ of a few times $10^4$ to $10^5$ would
be needed to produce significant direct synchrotron emission at
60~keV.  We can estimate the maximum Lorentz factor to which electrons
will be accelerated by setting the diffusive shock acceleration time
scale equal to the synchrotron cooling time scale.  Following the
approach of \citet{markoff01}, we equate the synchrotron loss rate to
a conservative acceleration rate \citep[see][]{jokipii87} to find a
maximum Lorentz factor to which electrons can be accelerated of
$\gamma_{max} \approx 2700 (B/10^4\,{\rm G})^{-1/2} (\xi /
100)^{-1/2}$, where $\xi$ is the ratio of the diffusive scattering
mean free path to the gyroradius.  Thus $\gamma_{max}$ is one to two
orders of magnitude below $\gamma_{sync}$.  Moreover, whereas
significant power from a distribution of synchrotron-emitting
relativistic electrons would also be expected at energies below
15~keV, the 16-100~keV X-ray luminosity of $\sim 10\,L_\odot$ during
2003 and 2004 is already close to the total energy budget available
from accretion.  Finally, to generate the X-ray spectral index
observed by INTEGRAL from direct synchrotron emission, one would need
a distribution of electrons that is much steeper than expected from
standard theories of particle acceleration in shocks
\citep{ellison90}.  It is therefore unlikely that the hard X-ray
emission from RT~Cru was due to direct synchrotron emission.

As an aside, we note that the production of significant synchrotron
emission at radio wavelengths would not require such high Lorentz
factors.  The ratio of synchrotron power to IC power from an electron
is equal to the ratio of the energy density in the magnetic field to
energy density in photons \citep{rlbook}.  Examining this ratio as a
function of distance from the WD, we expect a similar amount of power
from direct synchrotron and IC scattering near the surface of the WD,
where the B field could be strong ($B \sim 10^4$~G).  We therefore
suggest that the next time RT~Cru is in a powerlaw hard X-ray state
like the one detected by INTEGRAL in 2003-2004, radio observations be
performed to look for radio synchrotron emission.

Hard X-rays can be produced with much lower Lorentz factors through IC
scattering. For photons scattering off of a thermal distribution of
electrons, \cite{reynolds03} give a relation between the observed
photon index $\Gamma$ and the Compton $y$ parameter, which is related to
the factor by which the average photon energy increases.  A steep
photon index of $\Gamma > 2$ indicates a $y$ parameter less than 1, in
which case there is no significant up-scattering.  Therefore, since
RT~Cru had a steep photon index of $\Gamma \approx 2.7$, scattering
off of a thermal distribution of electrons could not have been
responsible for the powerlaw hard X-ray emission.  Non-thermal,
relativistic electrons must have been involved in producing the
powerlaw spectrum observed by INTEGRAL.  RT~Cru thus contains a white
dwarf that can accelerate electrons to relativistic speeds.  Three
other systems that contain WDs that also generate relativistic
electrons are CH~Cyg, RS~Oph, and possibly R~Aqr, all of which have
jets \citep{rupen07,obrien06,crocker2001,nichols07}.

Assuming now that the powerlaw hard X-ray spectrum was due to IC
scattering from a powerlaw distribution of relativistic electrons, we
can estimate the location of the scattering electrons.  Since the
strength of a dipole $B$ field falls like $1/r^3$, whereas the energy
density in the photon field only falls like $1/r^2$, IC scattering
will dominate as one moves farther from the WD.  If we ask how far
away from the radiation source the IC region should be to give us an
IC cooling time on the order of a year (the approximate duration of
the powerlaw hard X-ray state), we find that it should be a few tenths
of an AU away.  Given the orbital period for this systems, that could
put the IC emission region either between the WD and the red giant, as
in a colliding-winds region, above the WD disk in a corona, or at the
base of a jet. A model consisting of IC scattering off of relativistic
electrons in a corona or at the base of a jet reproduces the observed
hard X-ray emission well in X-ray binaries \citep{markoff05}.  The
steep powerlaw index, which is similar to that seen in the steep
powerlaw state in microquasars, could be due to a low scattering
optical depth \citep{rlbook}.

\section{Conclusions} \label{sec:conclusion}  

We have observed the first symbiotic star with X-ray emission out to
greater than 60~keV with the HETG on $Chandra$.  The stochastic
variability and cooling-flow type spectrum suggest that RT Cru is
powered by accretion onto a WD through a disk with an optically thin
boundary layer.  The accretion rate is near the top of the range for
which the boundary layer can remain optically thin.  The high initial
temperature of the cooling flow, and the high luminosity given the
accretion rate from the spectral fit suggest that the WD in RT~Cru
could be quite massive.  More generally speaking, it would be
difficult to get such hard X-ray emission from accretion onto a
low-mass WD with a shallow potential well and low Kepler velocity.
Given the nature of the powerlaw hard X-ray emission previously
observed by INTEGRAL, it therefore appears that the accreting,
non-magnetic WD in RT~Cru is able to generate a non-thermal, powerlaw
distribution of electrons and very hard X-ray emission through IC
scattering.  Three other systems in which WDs can accelerate electrons
to relativistic speeds \cite[RS~Oph, CH~Cyg, and possibly
R~Aqr;][]{rupen07,obrien06,crocker2001,nichols07} all have jets.
Radio observations of RT~Cru during the next powerlaw hard X-ray state
like that observed by INTEGRAL in 2003 and 2004 could play an
important role in diagnosing this system, and determining the extent
to which some symbiotic stars might constitute the nanoquasar analog
to microquasars \citep{zamanov02}.

\acknowledgments

We thank the referee, Marina Orio, for comments and suggestions which
improved the final quality of this article. We also thank K. Mukai,
F. Paerels, T. Maccarone, and A. J. Bird for useful discussions, R. Lopes de
Oliveira and D. Huenemoerder for help with the data analysis, and
S. Markoff for comments on the manuscript.  G.J.M. Luna acknowledges
support from CNPq (process 0141805/2003-0) and FAPESP (process
02/08816-5).  J.L.S. is supported by an NSF Astronomy and Astrophysics
Postdoctoral Fellowship under award AST-0302055.  Support for this
work was provided by the National Aeronautics and Space Administration
through Chandra Award Numbers DD5-6034X and NNX06AI16G issued by the
Chandra X-ray Observatory Center, which is operated by the Smithsonian
Astrophysical Observatory for and on behalf of the National
Aeronautics Space Administration under contract NAS8-03060.  We
acknowledge with thanks the variable star observations from the AAVSO
International Database contributed by observers worldwide and used in
this research.

\clearpage


\begin{deluxetable}{ll} 
\tabletypesize{\footnotesize}
\tablecolumns{2} 
\tablewidth{0pt}
\tablecaption{Cooling-flow model parameters. \label{tab1}} 
\tablehead{\colhead{Parameter} & \colhead{Value (Min, Max)\tablenotemark{a}}}
\startdata
$\dot{M}$\tablenotemark{b} ($10^{-9}$ M$_{\odot}$~yr$^{-1}$) & 1.8 (1.6, 2.0) \\ 
$kT_{max}$ (keV) & 80 (56,\nodata) \\
$n_H$: Full covering (10$^{22}\,{\rm cm}^{-2}$) & 8.2 (7.7, 8.8) \\
$n_H$: Partial covering (10$^{22}\,{\rm cm}^{-2}$) & 65 (52, 78) \\
Covering Fraction & 0.74 (0.69, 0.79) \\
Abundance (w.r.t. Solar\tablenotemark{c}) & 0.30 (0.02, 0.45) \\
$F_X$\tablenotemark{d} ($10^{-12}$ erg cm$^{-2}$ s$^{-1}$) & 9.1 (7.5, 10.2) \\
$L_X$\tablenotemark{d} ($10^{34}$ erg s$^{-1}$) & 3.1 (2.4, 3.9) \\
\enddata
\tablenotetext{a}{90\% confidence upper and lower limits.}
\tablenotetext{b}{Accretion rate onto the compact object.}
\tablenotetext{c}{Using solar abundance of \cite{abund}.}
\tablenotetext{d}{$F_X$ is the absorbed 0.3--8.0~keV flux, and $L_X$ is
the unabsorbed 0.3 -- 8.0~keV luminosity ($d=2$~kpc).}
\tablecomments{The model consists of optically thin thermal
  emission from an isobaric cooling flow with absorbers that both
  fully cover and partially cover the source, plus a Gaussian
  line.}
\end{deluxetable}
\clearpage

\begin{deluxetable}{lccc}
\tabletypesize{\footnotesize}
\tablecolumns{4} 
\tablewidth{0pt}
\tablecaption{Iron lines. \label{tab2}} 
\tablehead{
\colhead{} & 
\colhead{Fe XXV } & 
\colhead{Fe XXVI} & 
\colhead{Fe K$\alpha$} 
}
\startdata
Line center (keV)\tablenotemark{a}: &
6.946$^{6.959}_{6.933}$&6.693$^{6.706}_{6.677}$
&6.379$^{6.399}_{6.358}$ \\  
EW\tablenotemark{b} (eV): & 72& 60& 108\\ 
\enddata
\tablenotetext{a}{Super- and subscripts represent 90\% confidence
  upper and lower limits, respectively.} 
\tablenotetext{b}{Gaussian-fit equivalent widths.  EW uncertainties
  are on the order of 10--15\%.} 
\end{deluxetable}
\clearpage

\clearpage

\end{document}